# Unconventional Superconducting Phase Diagram of Monolayer WTe$_2$

Tiancheng Song[1,2], Yanyu Jia[1], Guo Yu[1,3], Yue Tang[1], Ayelet J. Uzan[1], Zhaoyi Joy Zheng[1,3], Haosen Guan[1], Michael Onyszczak[1], Ratnadwip Singha[4], Xin Gui[4,5], Kenji Watanabe[6], Takashi Taniguchi[7], Robert J. Cava[4], Leslie M. Schoop[4], N. P. Ong[1], Sanfeng Wu[1*]

[1]Department of Physics, Princeton University, Princeton, New Jersey 08544, USA.
[2]Department of Physics, University of Wisconsin-Madison, Madison, Wisconsin 53706, USA.
[3]Department of Electrical and Computer Engineering, Princeton University, Princeton, New Jersey 08544, USA.
[4]Department of Chemistry, Princeton University, Princeton, New Jersey 08544, USA.
[5]Department of Chemistry, University of Pittsburgh, Pittsburgh, Pennsylvania 15260, USA
[6]Research Center for Electronic and Optical Materials, National Institute for Materials Science, 1-1 Namiki, Tsukuba 305-0044, Japan.
[7]Research Center for Materials Nanoarchitectonics, National Institute for Materials Science, 1-1 Namiki, Tsukuba 305-0044, Japan.

*Email: sanfengw@princeton.edu

The existence of a quantum critical point (QCP) and fluctuations around it are believed to be important for understanding the phase diagram in unconventional superconductors such as cuprates, iron pnictides, and heavy fermion superconductors. However, the QCP is usually buried deep within the superconducting dome and is difficult to investigate. The connection between quantum critical fluctuations and superconductivity remains an outstanding problem in condensed matter. Here combining both electrical transport and Nernst experiments, we explicitly demonstrate the onset of superconductivity at an unconventional QCP in gate-tuned monolayer tungsten ditelluride (WTe$_2$), with features incompatible with the conventional Bardeen–Cooper–Schrieffer (BCS) scenario. The results lead to a novel superconducting phase diagram that is distinguished from other known superconductors. Two distinct gate-tuned quantum phase transitions are observed at the ends of the superconducting dome. We find that quantum fluctuations around the QCP of the underdoped regime are essential for understanding how the monolayer superconductivity is established. The unconventional phase diagram we report here illustrates a previously unknown relation between superconductivity and QCP.

## I. INTRODUCTION

While the application of general arguments based on symmetry, topology, and strong correlations have greatly advanced our knowledge of quantum phases of matter, much remains to be understood regarding the role of quantum fluctuations. At zero temperature, a quantum phase transition (QPT) occurs when quantum fluctuations are large enough to destroy a phase with long-range order [1–3]. The experimental characterization of a QPT and the associated quantum critical point (QCP) is a challenging task. Notable examples are QPTs in superconductors [4–6]. While the Landau-Ginzburg theory provides a general framework of superconducting transitions in conventional bulk superconductors, phase transitions in two-dimensional (2D) thin films are described by the Berezinskii–Kosterlitz–Thouless (BKT) theory, in which unbinding of 2D vortices and antivortices plays a critical role. However, the BKT theory [7] usually describes a transition driven by thermal rather than quantum fluctuations. Despite extensive efforts and progress, the understanding of QPTs from a superconducting to a resistive state in 2D, observed in various systems [4–6] tuned by, e.g., magnetic field, disorder strength and electron doping (**Fig. 1a**), remains unsatisfactory (see, for example, a recent review [5]).

In addition to 2D superconductors, the physics of a QCP could be important to understand the phase diagram and superconductivity in several unconventional superconductors, such as cuprates [8], iron pnictides [9] and heavy fermion materials [10]. In these cases, quantum fluctuations near a QCP, at which the nearby ordered phase is no longer stable, are believed to play a key role in defining the superconducting dome centered at the QCP (**Fig. 1b**). The QCPs here are typically buried in the dome [8–10], which hampers experiments. In general, the underlying physics of QCPs and the connection to the unconventional superconductivity in these intriguing situations remain to be understood.

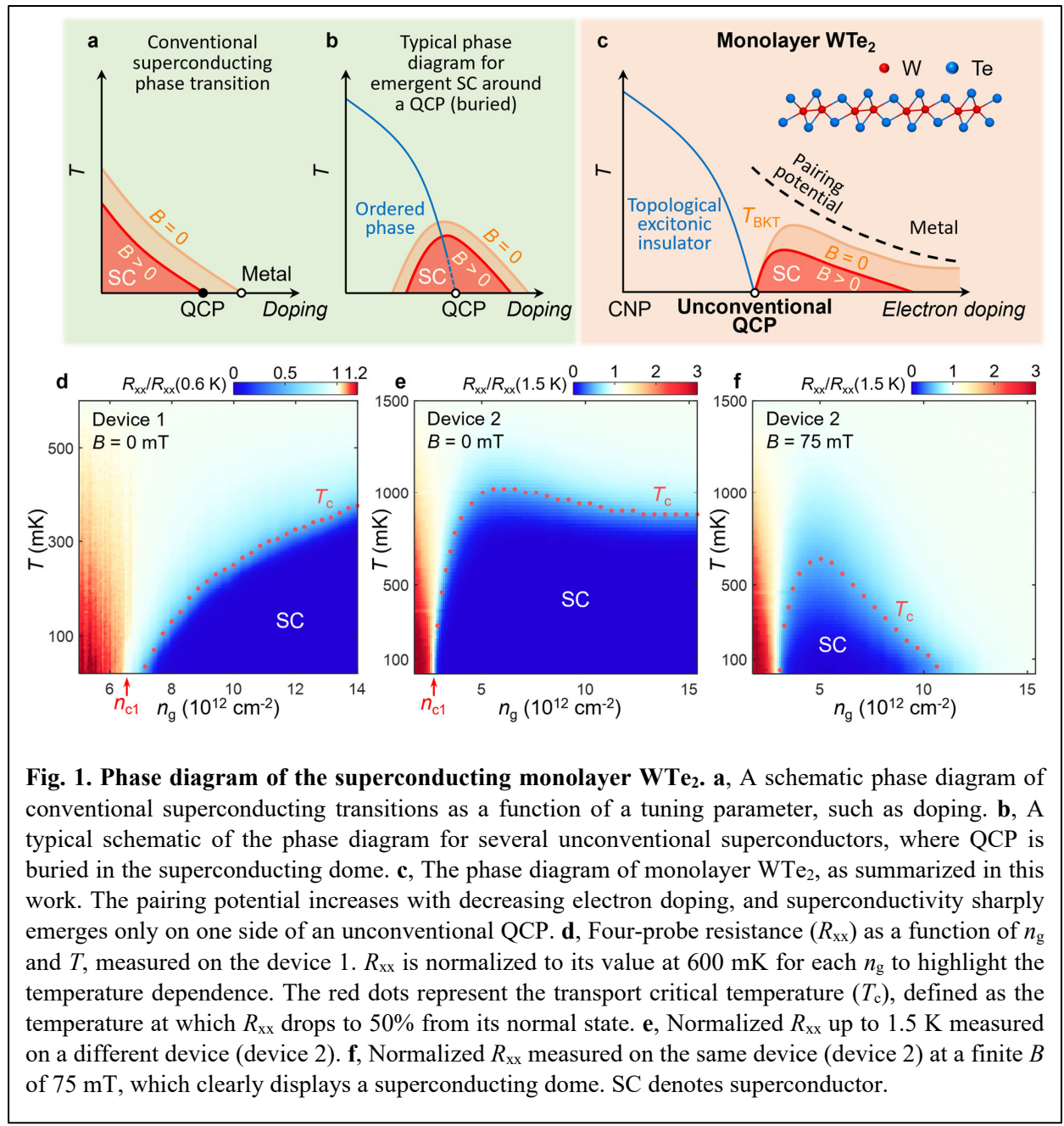

**Fig. 1. Phase diagram of the superconducting monolayer WTe$_2$. a**, A schematic phase diagram of conventional superconducting transitions as a function of a tuning parameter, such as doping. **b**, A typical schematic of the phase diagram for several unconventional superconductors, where QCP is buried in the superconducting dome. **c**, The phase diagram of monolayer WTe$_2$, as summarized in this work. The pairing potential increases with decreasing electron doping, and superconductivity sharply emerges only on one side of an unconventional QCP. **d**, Four-probe resistance ($R_{xx}$) as a function of $n_g$ and $T$, measured on the device 1. $R_{xx}$ is normalized to its value at 600 mK for each $n_g$ to highlight the temperature dependence. The red dots represent the transport critical temperature ($T_c$), defined as the temperature at which $R_{xx}$ drops to 50% from its normal state. **e**, Normalized $R_{xx}$ up to 1.5 K measured on a different device (device 2). **f**, Normalized $R_{xx}$ measured on the same device (device 2) at a finite $B$ of 75 mT, which clearly displays a superconducting dome. SC denotes superconductor.

In this work, based on a series of findings in monolayer tungsten ditelluride (WTe$_2$), we report an unconventional phase diagram that reveals a new relation between a QCP and superconductivity. We summarize the key result -- an unconventional phase diagram of monolayer WTe$_2$, in **Fig. 1c.** In varying the gate-induced electron doping ($n_g$), we have uncovered the existence of a superconducting dome that was not apparent in previous experiments [11–13]. A recent vortex Nernst experiment [13] provided a hint of its existence from the steep decrease in the pairing potential with increasing $n_g$. The direct observation of the superconducting dome here allows us to directly access two $n_g$-tuned QPTs from the superconducting state in the same device: one defines the upper end of the dome in the overdoped regime, while the other defines the lower end in the underdoped regime. The overdoped QPT appears to follow conventional expectations, but the underdoped QPT features an unconventional QCP at the doping $n_{c1}$ (identified in Ref. 13). Remarkably, as we change the magnetic field ($B$), the latter remains fixed. At large $B$, superconductivity emerges at a singular value of $n_g$ right above $n_{c1}$. As $B$ is lowered to zero, this point spawns the entire regime of $n_g > n_{c1}$, but the superconductivity is strictly forbidden at $n_g < n_{c1}$. Using electrical transport and Nernst measurements, we report and investigate this highly asymmetric superconducting dome on one side of the QCP.

## II. RESULTS

### A. Superconducting dome of the monolayer WTe$_2$

The monolayer WTe$_2$ devices and measurement schemes used in this work follow our earlier report [13]. **Figure 1d** plots the electronic phase diagram of the monolayer WTe$_2$ observed in the four-probe resistance ($R_{xx}$) measured in device 1. It displays a critical electron density ($n_{c1}$) of ~ $6.5 \times 10^{12}$ cm$^{-2}$ and a monotonically increasing transport critical temperature ($T_c$) when the carrier density ($n_g$) increases, consistent with previous results [11–13]. The vortex Nernst experiment [13] revealed that $R_{xx}$ of a 2D superconductor is sensitive to the BKT transition (at $T_{BKT}$) but not the pairing potential (or the superconducting gap, $\Delta$). Interestingly the pairing potential of monolayer WTe$_2$ is stronger at lower $n_g$ [13], opposite to the trend of $T_c$ v.s. $n_g$. The usual Bardeen–Cooper–Schrieffer (BCS) prediction of $\Delta \sim 1.76 k_B T_c$ is thus no longer valid here. The exact form of $T_c$ as a function of $n_g$ is strongly influenced by vortex pinning and hence influenced by disorder in the sample. To demonstrate this, we show the phase diagram of a different device (device 2) in **Fig. 1e**, fabricated following the same procedure (see **Fig. S1** for a comparison of the two devices). While the appearance of gate-tuned superconductivity is qualitatively consistent with device 1, $n_{c1}$ is now a much lower value, $2.7 \times 10^{12}$ cm$^{-2}$, and more importantly, $T_c$ is no longer a monotonic function of $n_g$. Instead, with increasing $n_g$ above $n_{c1}$, $T_c$ first rapidly increases to about 1 K and then decreases. We attribute the different transition temperatures observed in these devices to a varied amount of disorders and impurities, to which $T_c$ appears to be very sensitive. In our current fabrication process, we aim to minimize the level of disorders but they are still present. We hope a future generation of device fabrication that can better control and distinguish the types and amount of disorders could reveal more information on how $T_c$ is influenced by disorders.

The non-monotonic variation of $T_c$ suggests the presence of a dome-shaped superconducting regime if $n_g$ can be further increased. Unfortunately, in our typical devices, the high-$n_g$ end of the dome at zero $B$ requires a large $n_g$ that is beyond the limit of gating due to the dielectric breakdown of hexagonal boron nitride. However, with a finite $B$ (applied in the direction normal to the 2D flake), the superconductivity at high $n_g$ is suppressed first and then the superconducting dome becomes apparent (**Fig. 1f**). These observations unambiguously confirm that the pairing potential is indeed weaker at higher $n_g$. Based on the direct observation of the dome, we next uncover insights into the origin of the monolayer superconductivity.

## B. Onset of superconductivity at the QCP

**Figures 2a-i** present the resistance phase diagram spanned by $T$ and $n_g$, taken from device 2 in various $B$. At $B > 350$ mT, there is no hint of superconductivity (blue region) in the $R_{xx}$ diagram. When $B$ is lowered below ~ 250 mT, superconductivity emerges at the singular point $n_{c1}$. As $B$ further decreases to zero, the superconducting region rapidly expands to define the dome. Strikingly, the growth of the dome - strictly confined to the region above $n_g$ - is asymmetric to the extreme.

The sharp asymmetry also applies to fluctuations of the order parameter detected by the vortex Nernst signal (**Fig. 2j**). The applied temperature gradient $-\nabla T$ drives a current of mobile vortices which generates a Nernst voltage $V_N$ engendered by phase slippage of the condensate order parameter. The Nernst experiment provides a sensitive probe of superconducting fluctuations that cannot be seen in resistance measurements [13,14]. As reported [13], we derive maps of $V_N$ (**Figs. 2k-r**) under conditions corresponding to the $R_{xx}$ maps of **Figs. 2b-i**, respectively. One finds that at low $B$ no superconducting fluctuation is observed inside the dome (**Fig. 2r**), whereas, at the transition (i.e., the boundary of the dome), a strong Nernst signal develops. With increasing $B$, the dome shrinks, accompanied by strong fluctuations visible on the map. At high $B$ (e.g., at 250 mT, **Fig. 2m**), the fluctuations are concentrated near $n_{c1}$ (more precisely slightly above $n_{c1}$, i.e., $n_g \to n_{c1}^+$). The $V_N$ signal is observable at fields as high as 450 mT (**Fig. 2k**), well above the magnetic field needed to fully suppress the resistance signature of superconductivity in the $R_{xx}$ maps. The first appearance of superconducting fluctuations at $n_g \to n_{c1}^+$, confirmed in both $R_{xx}$ and $V_N$, implies the onset of superconductivity at $n_{c1}$. The superconducting fluctuations are large above the critical point at $n_{c1}$, but unresolvable below.

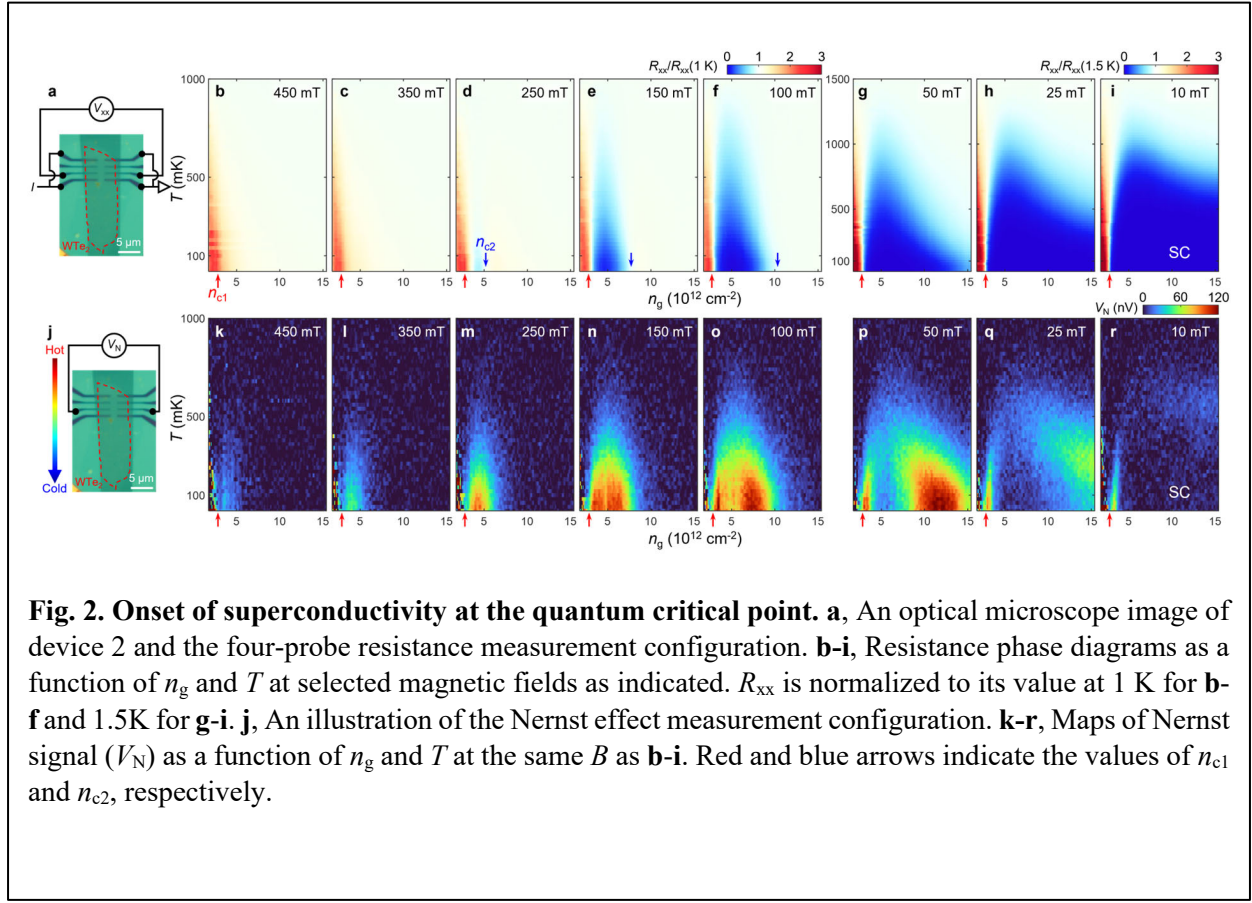

**Fig. 2. Onset of superconductivity at the quantum critical point. a**, An optical microscope image of device 2 and the four-probe resistance measurement configuration. **b-i**, Resistance phase diagrams as a function of $n_g$ and $T$ at selected magnetic fields as indicated. $R_{xx}$ is normalized to its value at 1 K for **b-f** and 1.5K for **g-i**. **j**, An illustration of the Nernst effect measurement configuration. **k-r**, Maps of Nernst signal ($V_N$) as a function of $n_g$ and $T$ at the same $B$ as **b-i**. Red and blue arrows indicate the values of $n_{c1}$ and $n_{c2}$, respectively.

## C. Two distinct QPTs

As mentioned, a striking feature in **Fig. 2** is that, whereas the superconducting region expands rapidly to the high-$n_g$ end as $B$ is lowered to zero, it never expands to the low-density region ($n_g < n_{c1}$). The Nernst signal (i.e., superconducting fluctuations) is strictly absent below $n_{c1}$; this is the sudden death phenomenon of the fluctuations pointed out earlier [13]. With the dome-shaped superconductivity, we now can directly access two independent gate-tuned superconducting QPTs in one device, induced by either decreasing or increasing $n_g$ respectively from the same superconducting state. The sudden death feature only occurs at the underdoped QPT (at $n_{c1}$), but not at the QPT in the overdoped region (at $n_{c2}$). This aspect is evident from the sharply distinct fluctuation patterns around the two QPTs shown in **Fig. 2r**. Whereas $V_N$ in the vicinity of the overdoped QPT extends over a very broad region, it is narrowly confined to the critical point in the underdoped QPT at $n_{c1}$. In supplementary **Fig. S2**, we also show that the two transitions are independent of the displacement electric field applied by the two gates.

We further emphasize this distinction in **Fig. 3**. The two QPTs can be clearly observed in $R_{xx}$ tuned by $n_g$, measured at the base $T$ (**Fig. 3a**). If $B$ is fixed at ~100 mT, $R_{xx}$ vanishes abruptly at $n_{c1}$ once it enters the superconducting state. This is followed by a subsequent increase in $R_{xx}$ when $n_g$ exceeds $n_{c2}$. A distinctive feature is that $n_{c2}$ is field sensitive, unlike $n_{c1}$ ($n_{c2} \sim 10.4 \times 10^{12}$ cm$^{-2}$ at $B \sim 100$ mT). The corresponding

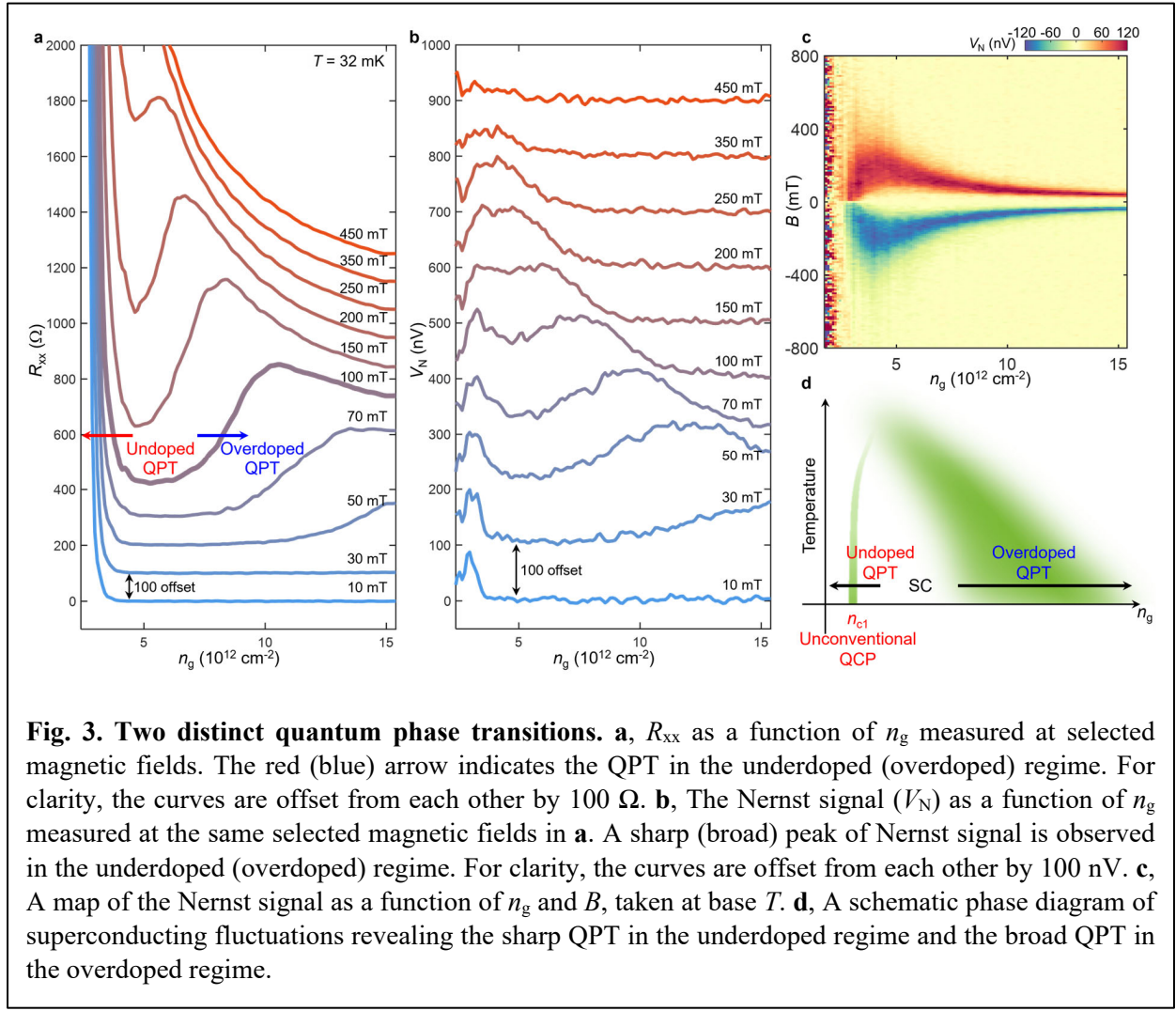

**Fig. 3. Two distinct quantum phase transitions. a**, $R_{xx}$ as a function of $n_g$ measured at selected magnetic fields. The red (blue) arrow indicates the QPT in the underdoped (overdoped) regime. For clarity, the curves are offset from each other by 100 Ω. **b**, The Nernst signal ($V_N$) as a function of $n_g$ measured at the same selected magnetic fields in **a**. A sharp (broad) peak of Nernst signal is observed in the underdoped (overdoped) regime. For clarity, the curves are offset from each other by 100 nV. **c**, A map of the Nernst signal as a function of $n_g$ and $B$, taken at base $T$. **d**, A schematic phase diagram of superconducting fluctuations revealing the sharp QPT in the underdoped regime and the broad QPT in the overdoped regime.

Nernst signals measured at fixed $B$ are displayed in **Fig. 3b**, as well as in the color map in **Fig. 3c**. The data confirms that $n_{c1}$ remains fixed as $B$ is varied, but $n_{c2}$ is highly sensitive to $B$. The overdoped QPT occurs over a much broader density range with a fluctuation tail that extends deep into the normal state. We summarize in **Fig. 3d** the fluctuation pattern inferred from the Nernst data (extrapolated to the zero $B$ limit). We conclude that the overdoped QPT is closer to conventional 2D superconductor-to-metal/insulator transitions observed in other systems including superconducting thin films [4–6] (as illustrated in **Fig. 1a**), whereas the underdoped QPT at $n_{c1}$ is unconventional.

### D. Unconventional pairing mechanism

An interesting feature of the superconductivity in monolayer $WTe_2$ is that it is unrelated – and unconnected – to a 3D parent superconducting state ($WTe_2$ in the bulk is not a superconductor). This distinction is shared with the superconductivity in magic-angle twisted graphene [15]. The fact that next to the superconducting state there is a topological insulator phase at charge neutrality suggests that the pairing nature of the superconductivity deserves careful studies. Topological aspects of unconventional superconductivity in monolayer $WTe_2$ are discussed in e.g. Ref [16,17].

Our results provide direct experimental evidence for an unconventional pairing mechanism in monolayer $WTe_2$. (*i*) The fact that the pairing potential increases with decreasing $n_g$ [13] suggests that electron interaction is important for pair formation. With lower $n_g$, the screening effect is reduced and hence stronger electron interactions. (*ii*) The abrupt appearance of superconductivity at $n_{c1}$ is incompatible with the phonon-mediated BCS scenario. In this scenario, the BCS gap [18] is $\Delta \sim \hbar\omega_D \exp(-1/NV)$, where $N$ is the density of state (DOS) near the Fermi energy, $V$ is the electron-phonon coupling strength, $\omega_D$ is the Debye cutoff frequency, and $\hbar$ is the Planck constant. For a small window of $n_g$ near $n_{c1}$, we do not expect a dramatic change in $V$ or $\omega_D$. Thus in the BCS framework, one would need to assign the onset of superconductivity near $n_{c1}$ (**Fig. 2d**) to a dramatic enhancement of $N$, such as the presence of a van Hove singularity (vHS), near this specific doping. We next demonstrate that this is not the case here.

**Figures 4a & b** display the $R_{xx}$ maps tuned by $B$ and $n_g$, taken at base $T$ and 1 K respectively, both again demonstrating the onset of superconductivity right above $n_{c1}$. **Figure 4c** plots the four-probe conductance ($G_{xx} \equiv 1/R_{xx}$) as a function of $n_g$ at selected $B$, where the peaks correspond to the growth of superconducting regions. One finds that when superconductivity is fully suppressed at high $B$, $G_{xx}$ depends on $n_g$ linearly as expected from the Drude model. The same linear relation can be found at high $T$ (**Fig. S3**). Interestingly, the conductance peak of superconductivity emerges in the middle of the linear regime (at $n_p$ as indicated in **Fig. 4d** upper panel), where no features in $G_{xx}$ are observed, indicating the absence of vHS or other types of anomalous enhancement in the DOS at this doping. In addition, the presence of vHS would also lead to a sign change in the Hall signal. In **Fig. 4d** (lower panel) and **Fig. S4**, we present the Hall effect measurements which reveal a nearly linear dependence of the Hall density *v.s.* $n_g$. Clearly no sign change is observed near the onset of superconductivity. The Hall data again confirms the absence of any anomalous features in the DOS near $n_{c1}$ in the normal state. Similar behaviors are consistently observed in device 1 (**Fig. S5**), where $n_{c1}$ is at a much higher density. The fact that $n_{c1}$ depends sensitively on the sample details (e.g., disorder strength) may be regarded as another manifestation of unconventional superconductivity since the conventional BCS gap is expected to be robust against disorder as per Anderson's argument [19]. We conclude that the sharp onset of superconductivity at $n_{c1}$ in monolayer $WTe_2$ cannot be attributed to the phonon-mediated BCS mechanism; instead, it suggests that superconductivity stems from the quantum fluctuations near the unconventional QCP at $n_{c1}$, where a pre-existing ordered phase below $n_{c1}$ is suppressed. This ordered phase could be the topological excitonic insulator state [20–28]. Unlike other unconventional superconductors such as high-$T_c$ and heavy fermion systems [8–10], in the case of monolayer $WTe_2$, the

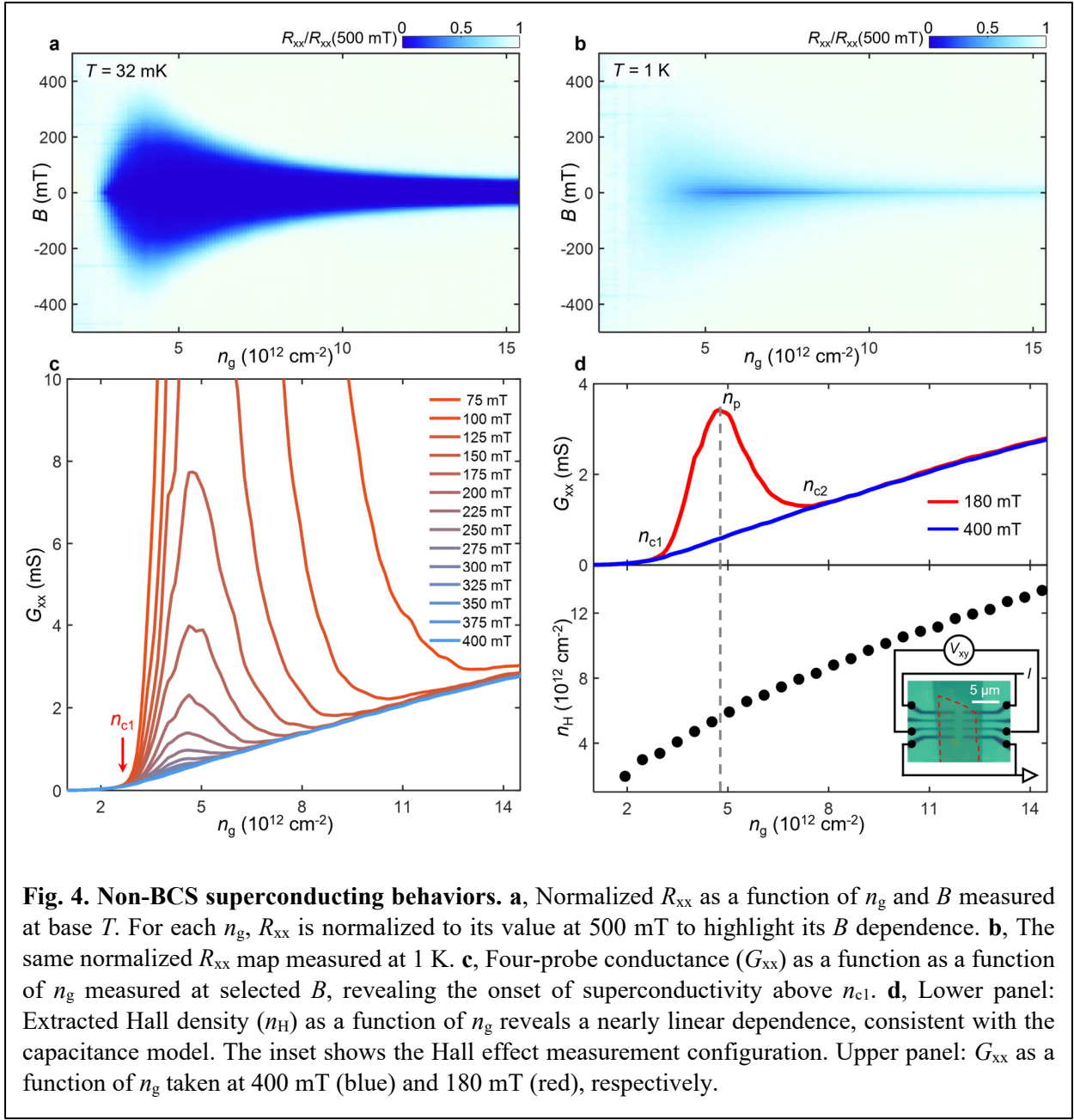

**Fig. 4. Non-BCS superconducting behaviors. a**, Normalized $R_{xx}$ as a function of $n_g$ and $B$ measured at base $T$. For each $n_g$, $R_{xx}$ is normalized to its value at 500 mT to highlight its $B$ dependence. **b**, The same normalized $R_{xx}$ map measured at 1 K. **c**, Four-probe conductance ($G_{xx}$) as a function as a function of $n_g$ measured at selected $B$, revealing the onset of superconductivity above $n_{c1}$. **d**, Lower panel: Extracted Hall density ($n_H$) as a function of $n_g$ reveals a nearly linear dependence, consistent with the capacitance model. The inset shows the Hall effect measurement configuration. Upper panel: $G_{xx}$ as a function of $n_g$ taken at 400 mT (blue) and 180 mT (red), respectively.

fluctuations near the QCP only give rise to superconductivity on one side of the QCP ($n_g > n_{c1}$), whereas superconductivity is strictly forbidden on the other side.

### III. Summary

In this work, we provide new insights into the origin of the surprising superconductivity in monolayer WTe$_2$, a state arising from a doped excitonic quantum spin Hall insulator [20–28]. The results reveal a new relation between superconductivity and quantum criticality not seen previously. The asymmetry suggests a novel type of quantum critical phenomena. Theoretically, a novel deconfined QCP [29–32] has been proposed in a model describing a transition between a quantum spin Hall insulator and a superconductor. The experimental phenomenology observed here is consistent with the expectation of a deconfined QCP, although further studies are needed to clarify their connections and establish a concrete theory behind the phenomena.

## ACKNOWLEDGEMENTS

This work is mainly supported by AFOSR Young Investigator Program (FA9550-23-1-0140) to S.W. Materials synthesis and device fabrication are partially supported by the Materials Research Science and Engineering Center (MRSEC) program of the National Science Foundation (DMR-2011750) through support to R.J.C., L.M.S., N.P.O., and S.W. Data collection are partially supported by NSF through a CAREER award (DMR-1942942) to S.W. and ONR through a Young Investigator Award (N00014-21-1-2804) to S.W. N.P.O. is supported by the U.S. Department of Energy (DE-SC0017863). S.W., L.M.S., and N.P.O. acknowledge support from the Gordon and Betty Moore Foundation through Grants GBMF11946, GBMF9064, and GBMF9466, respectively. L.M.S is also supported by the David and Lucile Packard Foundation. T.S. acknowledges the Princeton Physics Dicke Fellowship program and is supported by the Office of the Vice Chancellor for Research and Graduate Education at the University of Wisconsin–Madison with funding from the Wisconsin Alumni Research Foundation. A.J.U. acknowledges support from the Rothschild Foundation and the Zuckerman Foundation. K.W. and T.T. acknowledge support from the JSPS KAKENHI (Grant Numbers 20H00354 and 23H02052) and World Premier International Research Center Initiative (WPI), MEXT, Japan.

# Supplementary Materials for

## Unconventional Superconducting Phase Diagram of Monolayer WTe$_2$

Tiancheng Song[1,2], Yanyu Jia[1], Guo Yu[1,3], Yue Tang[1], Ayelet J. Uzan[1], Zhaoyi Joy Zheng[1,3], Haosen Guan[1], Michael Onyszczak[1], Ratnadwip Singha[4], Xin Gui[4,5], Kenji Watanabe[6], Takashi Taniguchi[7], Robert J. Cava[4], Leslie M. Schoop[4], N. P. Ong[1], Sanfeng Wu[1*]

[1]Department of Physics, Princeton University, Princeton, New Jersey 08544, USA.
[2]Department of Physics, University of Wisconsin-Madison, Madison, Wisconsin 53706, USA.
[3]Department of Electrical and Computer Engineering, Princeton University, Princeton, New Jersey 08544, USA.
[4]Department of Chemistry, Princeton University, Princeton, New Jersey 08544, USA.
[5]Department of Chemistry, University of Pittsburgh, Pittsburgh, Pennsylvania 15260, USA
[6]Research Center for Electronic and Optical Materials, National Institute for Materials Science, 1-1 Namiki, Tsukuba 305-0044, Japan.
[7]Research Center for Materials Nanoarchitectonics, National Institute for Materials Science, 1-1 Namiki, Tsukuba 305-0044, Japan.

[*]Email: sanfengw@princeton.edu

**This file includes**

Device fabrication

Electrical transport measurements

Nernst measurements

Figures S1 to S5

**Device fabrication**

The devices used in this study are the same ones used in the previous report[13] in which we report the discovery of the QCP itself. The data, analysis and physics presented in this study are new. Details of the fabrication process can be found in reference[13]. Here we again describe key device details. Bottom: The hBN and graphite flakes were exfoliated onto 285 nm $SiO_2$/Si substrates and then characterized by optical and atomic force microscopy. Only atomically clean and uniform flakes were used. The bottom hBN/graphite stacks were fabricated using the standard dry transfer technique and then released on $SiO_2$/Si substrates with pre-patterned Ti/Au (5/60 nm) metal pads and alignment marks. Ti/Au (2/6 nm) electrodes and microheaters were patterned on top of the bottom stacks using standard electron beam lithography, followed by cold development, reactive ion etching and metal deposition. Similar steps were also followed to fabricate Ti/Au (5/60 nm) electrodes connecting the thin electrodes to the pre-patterned metal pads. Before the final assembly, the pre-patterned bottom stacks were tip-cleaned using the contact mode of an atomic force microscope. Top: The top graphite/hBN stacks were first fabricated using the same dry transfer technique. High-quality $WTe_2$ crystals were exfoliated onto 285 nm $SiO_2$/Si substrates in an Argon glovebox (oxygen and water concentration less than 0.1 ppm). For each device, one high-quality monolayer $WTe_2$ flake was identified and then aligned with the top stack which picks it up. The top stack was finally released on the bottom stack. The monolayer $WTe_2$ flakes were in contact with the bottom electrodes and were fully encapsulated.

**Electrical transport measurements**

The electrical transport measurements were performed in a dilution refrigerator equipped with a superconducting magnet and a base temperature of 20 mK. The four-probe resistance measurements were performed using the standard AC lock-in technique with a frequency of 13 Hz and an AC current excitation of 5 nA. The base electron temperature is calibrated to be 32 mK at the base fridge temperature.

**Nernst measurements**

The Nernst measurements were performed on the same devices using the same dilution refrigerator as the electrical transport measurements. The Nernst experiments followed closely to the previous report[13]. Two microheaters were fabricated close to the monolayer $WTe_2$ flake, each being a thin and narrow metal stripe (8 nm thick and 200 nm wide) with a low-temperature resistance of about 1 kΩ. We utilized the dual-heater measurement scheme where alternating current was applied to the two microheaters with a frequency ($\omega$) of 13 Hz and a 90° phase shift between each other. This produced an alternating temperature gradient at the frequency of $2\omega$, and the Nernst voltage ($V_N$) across the two probes was detected at the frequency of $2\omega$. The Nernst effect measurement configuration can be found in Fig. 2.

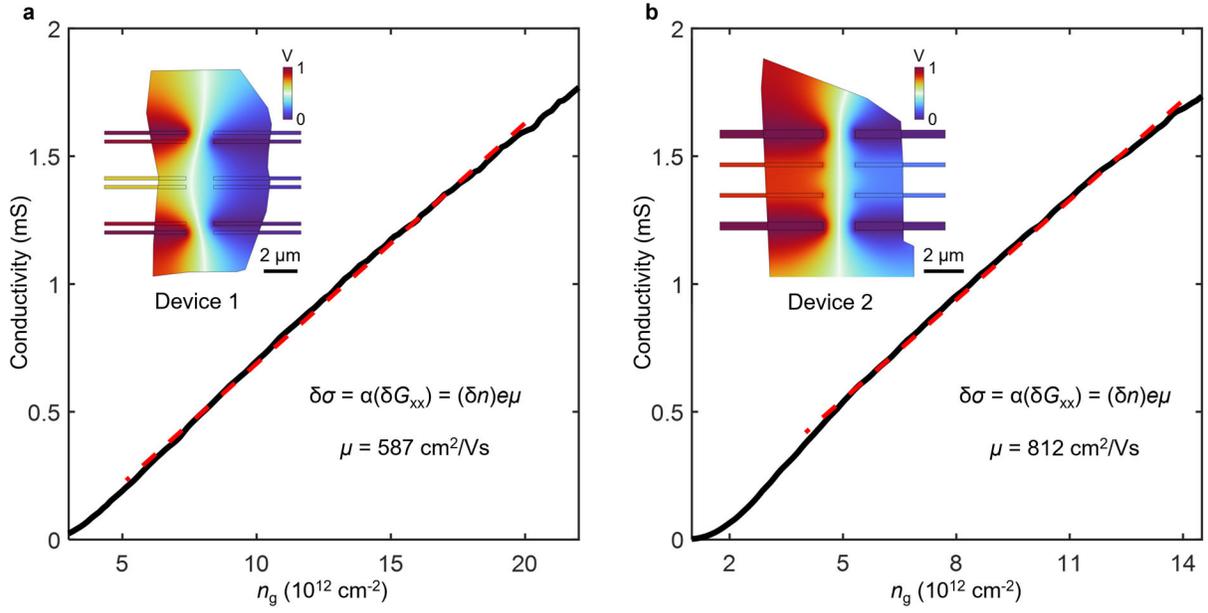

**Fig. S1. Estimation of electron mobility of the monolayer WTe$_2$ devices. a**, Conductivity as a function of $n_g$. Conductivity is calibrated using a geometry factor estimated by the electric potential simulation shown in the inset. The electron mobility is calculated based on the Drude model, $\delta\sigma = \alpha(\delta G_{xx}) = (\delta n)e\mu$, where $\sigma$ is the conductivity, $\alpha$ is the geometry factor, $n$ is the carrier density, $e$ is the electron charge, and $\mu$ is the electron mobility. The red dashed line shows the linear fit to extract the electron mobility of device 1, which is estimated to be 587 cm$^2$/Vs. **b**, Following the same procedure, the estimated electron mobility of device 2 is 812 cm$^2$/Vs.

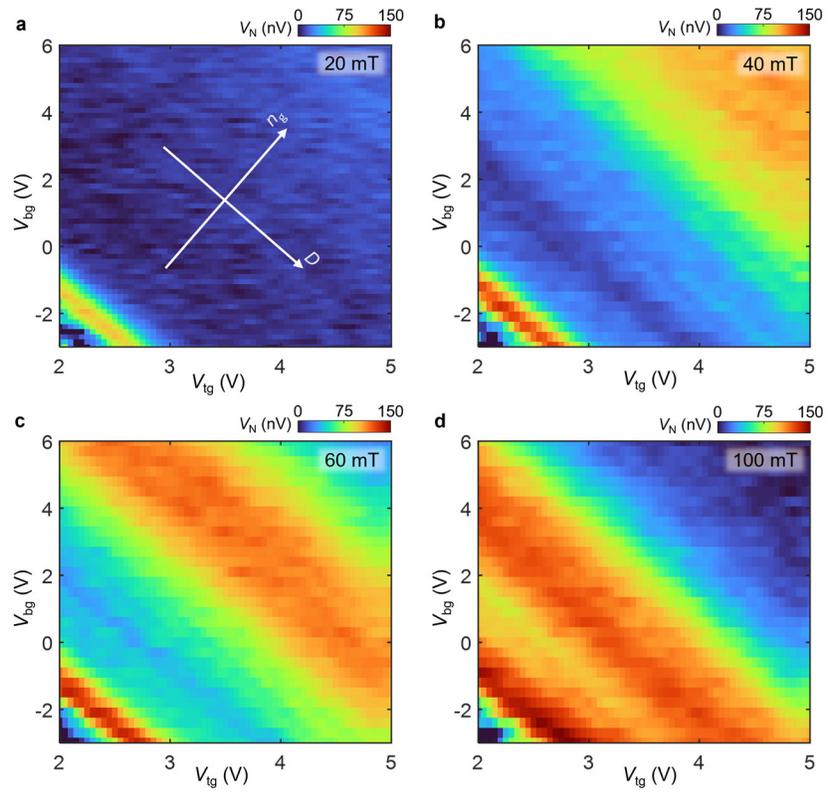

**Fig. S2. Nernst signal as a function of the top and bottom gate voltages measured at selected magnetic fields and base *T* from device 2.** The two white arrows indicate the tuning of carrier density and displacement field.

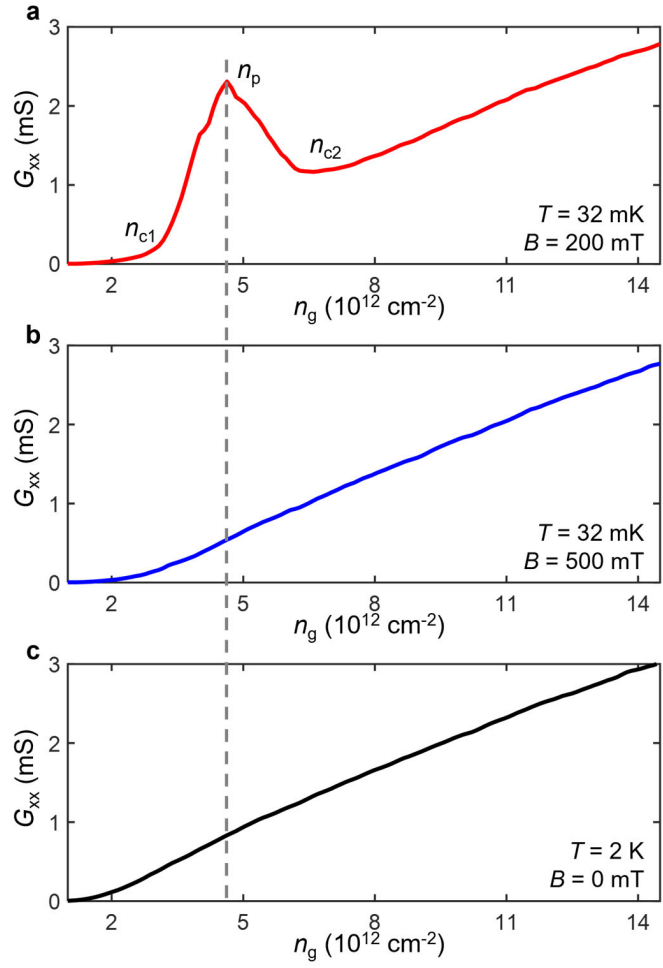

**Fig. S3. Additional Analysis on the birth of superconductivity in device 2. a**, $G_{xx}$ as a function of $n_g$ measured at based $T$ and a finite $B$ of 200 mT. The conductance peaks at $n_g \sim n_p$ in the range between $n_{c1}$ and $n_{c2}$. **b** and **c**, $G_{xx}$ as a function of $n_g$ measured for the $B$-induced normal state (at a high $B$ of 500 mT) (**b**) and the $T$-induced normal state (at 2 K) (**c**), revealing a nearly linear dependence on $n_g$.

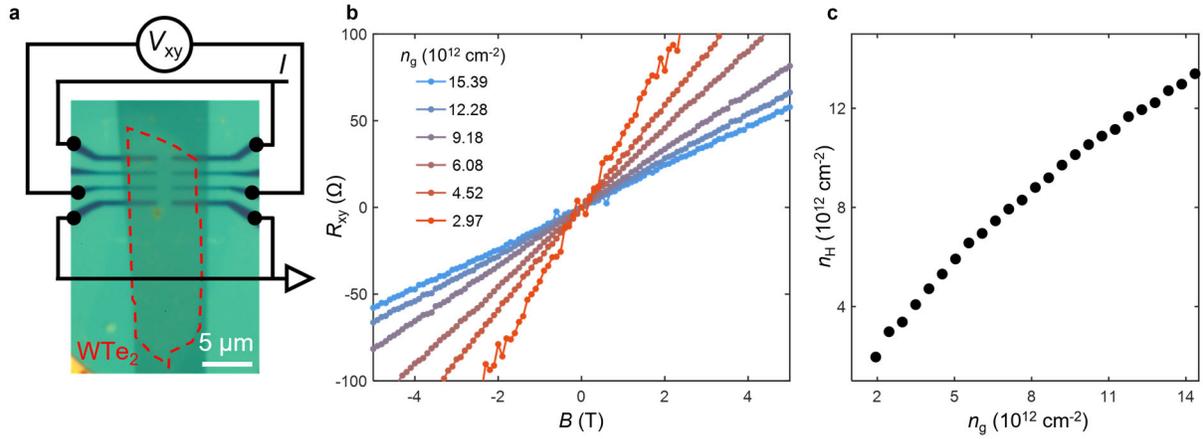

**Fig. S4. Hall effect measurements in device 2. a**, An optical microscope image of device 2 and the Hall effect measurement configuration. **b**, Hall resistance ($R_{xy}$) as a function of $B$ measured at selected carrier densities as indicated. A standard antisymmetrization process with respective to $B$ is applied to remove the mixing of $R_{xx}$ signals. **c**, Extracted Hall density ($n_H$) as a function of $n_g$, closely following the expectation from the capacitance model.

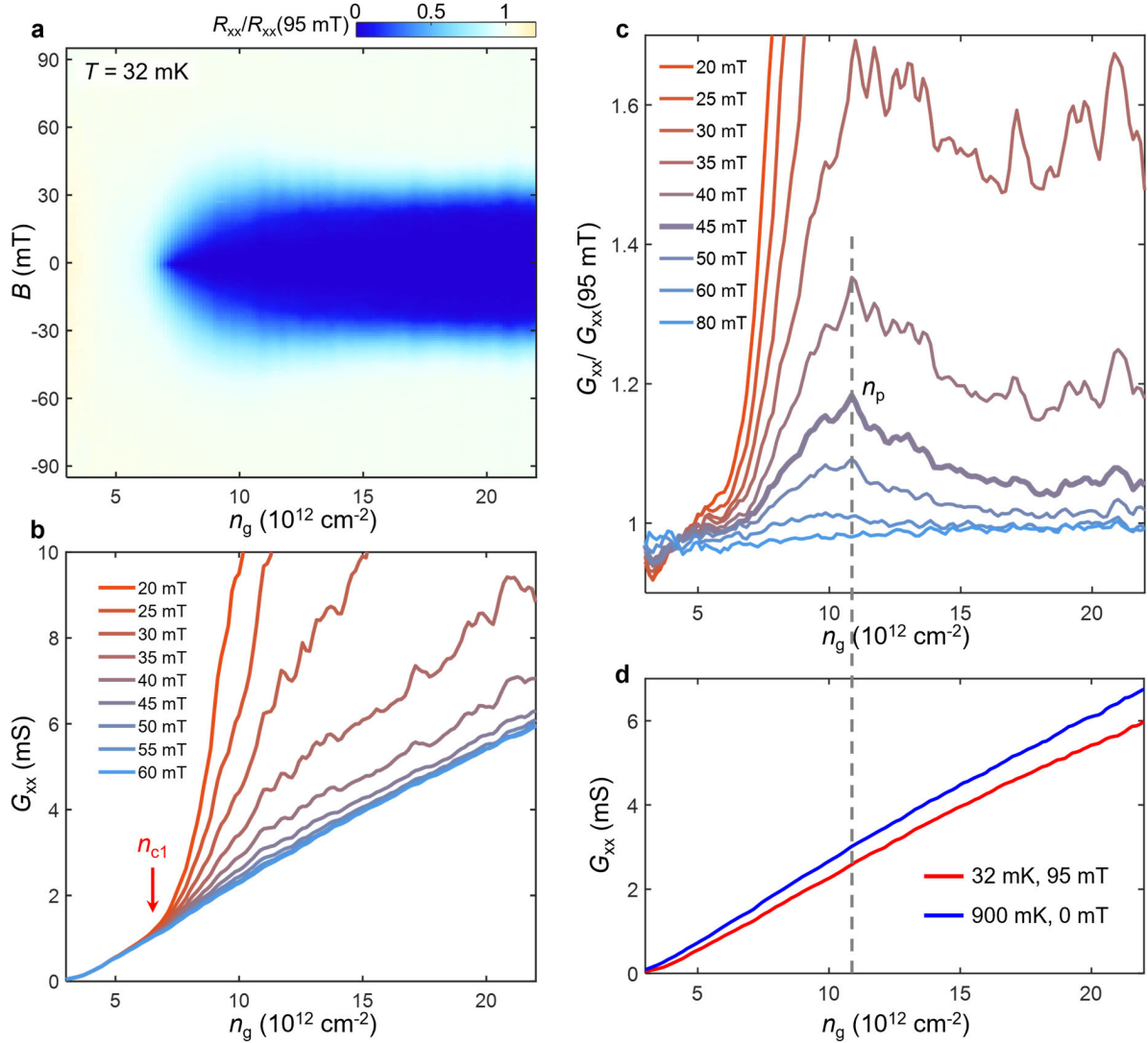

**Fig. S5. The birth of superconductivity in device 1. a**, Normalized $R_{xx}$ as a function of $n_g$ and $B$ measured at base $T$. For each $n_g$, $R_{xx}$ is normalized to its value at 95 mT to highlight its $B$ dependence. **b**, $G_{xx}$ as a function as a function of $n_g$ measured at selected $B$, revealing the onset of superconductivity above $n_{c1}$. **c**, Normalized $G_{xx}$ as a function of $n_g$ measured at selected $B$. For each $B$, $G_{xx}$ is normalized to its value at 95 mT to highlight its $n_g$ dependence. Near 45 mT, the conductance peak is clearly visible near $n_p$. **d**, The normal state $G_{xx}$ as a function of $n_g$ measured either at a high magnetic field of 95 mT (red) or a high temperature of 900 mK (blue), revealing a nearly linear dependence on $n_g$.

7